\documentclass[aps,pra,superscriptaddress,amsmath,amssymb,preprintnumbers,showpacs,twocolumn]{revtex4-1}
\usepackage{amssymb}
\usepackage{graphicx}
\usepackage{bm}
\usepackage{color}
\usepackage{subfigure}

\begin{document}

\title{Entropy production and information fluctuations along quantum trajectories}

\author{B. Leggio}
\affiliation{Dipartimento di Fisica e Chimica, Universit\`{a} di Palermo, Via Archirafi 36, 90123
Palermo, Italy}
\affiliation{Physikalisches Institut, Universit\"{a}t Freiburg,
Hermann-Herder-Stra\ss{}e 3, D-79104 Freiburg, Germany}

\author{A. Napoli}
\affiliation{Dipartimento di Fisica e Chimica, Universit\`{a} di Palermo, Via Archirafi 36, 90123
Palermo, Italy}

\author{A. Messina}
\affiliation{Dipartimento di Fisica e Chimica, Universit\`{a} di Palermo, Via Archirafi 36, 90123
Palermo, Italy}

\author{H.-P. Breuer}
\affiliation{Physikalisches Institut, Universit\"{a}t Freiburg,
Hermann-Herder-Stra\ss{}e 3, D-79104 Freiburg, Germany}

\newcommand{\ket}[1]{\displaystyle{|#1\rangle}}
\newcommand{\bra}[1]{\displaystyle{\langle #1|}}

\date{\today}

\begin{abstract}
Employing the stochastic wave function method, we study quantum features of
stochastic entropy production in nonequilibrium processes of open systems.
It is demonstrated that continuous measurements on the environment introduce
an additional, non-thermal contribution to the entropy flux, which is shown to be a
direct consequence of quantum fluctuations. These features lead to a quantum
definition of single trajectory entropy contributions, which accounts for the
difference between classical and quantum trajectories and results in a quantum
correction to the standard form of the integral fluctuation theorem.
\end{abstract}


\maketitle

\section{Introduction}
Fluctuation theorems (FTs) for nonequilibrium processes \cite{jarzrev, esporev} are a set of general laws describing the intrinsic fluctuating nature of thermodynamical
quantities for systems far from equilibrium. They describe the probability distribution of measurement outcomes for quantities such as energy or entropy. These laws for classical systems have been theoretically predicted \cite{jarz, crooks, espo, saga, garcia} and experimentally verified \cite{sciencejarz, seif1, seif2, genjarz} under various conditions, and a classical formulation of FTs has been satisfactory given and is nowadays an (almost) settled problem. On the other hand, many efforts to provide quantum versions of these laws have been made \cite{hanggirev, hanggi1, muka, kawa, kafri, lutz, liu}, but the quantum counterpart to FTs has not yet been fully understood. A theoretical description of stochastic entropy production has been given for classical \cite{seif} and quantum trajectories \cite{deff}, and employed for classical as well as quantum FTs \cite{muka, espo, kawa, seif1, seif2, lnbm}. These approaches have never taken into account the full quantum features of stochastic dynamics: When considering quantum systems and the probability distribution of measurement outcomes, the role of the external observer can not be neglected and a full quantum description of single nonequilibrium processes can not be given without incorporating the backaction of measurements \cite{wise1, wise2}. Following a quantum trajectory amounts indeed to a continuous measurement process, and it is reasonable to expect that it introduces a non-negligible term in entropy production. Moreover, because any measurement fundamentally affects the evolution of the monitored system, the previous proposals of quantum trajectories based on measurements on the open quantum system \cite{hanggi, deff, kafri} can not fully reproduce its quantum features, since these projective measurements partly hide the quantumness of the process. In the framework of open quantum dynamics, however, the time evolution of an open system can be monitored by continuous measurements on its environment \cite{wiseman}.

In this paper we apply the quantum stochastic wave function method \cite{carmichael,dal,zoller,plen} within the Markovian approximation to a generic quantum system interacting with a bath and, in general, externally driven through a fixed protocol, to obtain an expression for its entropy production. In this framework we consider information (entropy) contributions as extracted by measurements on the environment.
The ensemble Markovian dynamics of the open quantum system is described by the master equation
\begin{equation}\begin{split}\label{ME}
  \dot{\rho}&=-i\left[H_S(t),\rho\right]\\
  &+\sum_{i}\gamma_{i}(t)\bigg(A_{i}(t)\rho A_{i}^{\dag}(t)-\frac{1}{2}\Big\{A_{i}^{\dag}(t)A_{i}(t),\rho\Big\}\bigg),
\end{split}\end{equation}
with $\gamma_i(t)\geq0\,\,\,\,\,\forall i,t$. $H_S(t)$ is the free Hamiltonian of the open quantum system and the non-Hermitian operators $\{A_i\}$ are known as Lindblad operators. The time dependence of $H_S(t)$, $\gamma_i(t)$ and $A_i(t)$ originates from the open system-environment interaction and from the external protocol driving the open system out of a stationary state. Starting from this equation we will describe, in what follows, single realisations of the same nonequilibrium dynamics and introduce their associated entropy production.

The paper is structured as follows. In Sec.~\ref{stochwave} we introduce the stochastic wave function method and use it to define forward and backward quantum processes. Entropy contributions associated to them are introduced and discussed in Sec.~\ref{entropy}, and employed in Sec.~\ref{IFT} to derive a quantum correction to the standard form of integral fluctuation theorems. Our results are exemplified in Sec.~\ref{examp}, where some model systems are studied. Finally, we draw our conclusions in Sec.~\ref{conclu}.

\section{Stochastic wave function method and quantum nonequilibrium processes}\label{stochwave}
The stochastic wave function approach to quantum systems whose ensemble evolution is given by Eq.~(\ref{ME}) describes single realizations of the dissipative process by means of quantum trajectories representing the pure state evolutions of the open system which are conditioned on certain continuous measurements on the environment. The measurement events lead either to discontinuous, random transitions of the state vector of the open system (referred to as \textit{quantum jumps}) or to a continuous time evolution resulting from the no-jump events
(referred to as \textit{drift contribution}).

These pure state dynamical evolutions conditioned on the measurement outcomes are, mathematically speaking, piecewise deterministic processes (PDPs) \cite{breub} characterized by jumps described by the action of Lindblad operators $A_i$ introduced in Eq.~(\ref{ME}), each of which happens at a random time and along a randomly chosen channel $\{\gamma_{i_k},A_{i_k}\}$, and by a nonunitary deterministic time evolution between two jumps at $t_s$ and $t_f$, given by the effective time evolution operator $U_{\mathrm{eff}}(t_f,t_s)=\mathcal{T}\exp\left\{-i\int_{t_s}^{t_f}H_{\mathrm{eff}}(t)\mathrm{d}t\right\}$, where $H_{\mathrm{eff}}(t)=H_S(t)-\frac{i}{2}\sum_i\gamma_iA_i^{\dag}A_i$. A single quantum jump is therefore given by the transition
\begin{equation}\label{gjump}
|\chi\rangle\rightarrow|\psi_{i_k}\rangle=\frac{A_{i_k}|\chi\rangle}{||A_{i_k}|\chi\rangle||},
\end{equation}
while a drift is described by 
\begin{equation}\label{gdrift}
|\psi(t_f)\rangle=\frac{U_{\mathrm{eff}}(t_f,t_s)|\psi(t_s)\rangle}{||U_{\mathrm{eff}}(t_f,t_s)|\psi(t_s)\rangle||}.
\end{equation}
The normalisation of the state in Eqs.~\eqref{gjump} and~\eqref{gdrift} is necessary since the action of a Lindblad operator and of $U_{\mathrm{eff}}$ on a normalised states yields, in general, an unnormalised state.
Jumps occur with rates $\gamma_{i_k}||A_{i_k}|\psi\rangle||^2$. On the other hand the probability that, after jumping at time $t_k$, the system performs no further transitions up to time $t_{k+1}$ is $||U_{\mathrm{eff}}(t_{k+1},t_{k})|\psi\rangle||^2$.

Since the wave function plays here the role of a stochastic variable, at each time instant one associates to it a probability density $P[\psi,t]$. The meaning of such a density is that the product $P[\psi,t]\mathrm{d}\psi$ expresses the probability for the wave function of the system to lie, at time $t$, within the volume element $\mathrm{d}\psi$. Given any function $F[\psi]$ of the vector $|\psi\rangle$, its expectation value is evaluated as $E[ F[\psi]]=\int\mathrm{d}\psi F[\psi]P[\psi,t]$.
In particular, the density matrix is given by $\rho(t) = E[|\psi(t)\rangle\langle\psi(t)|]=\int\mathrm{d}\psi |\psi\rangle\langle\psi| P[\psi,t]$.

\begin{center}
\begin{figure}[h!]
\includegraphics[width=250pt]{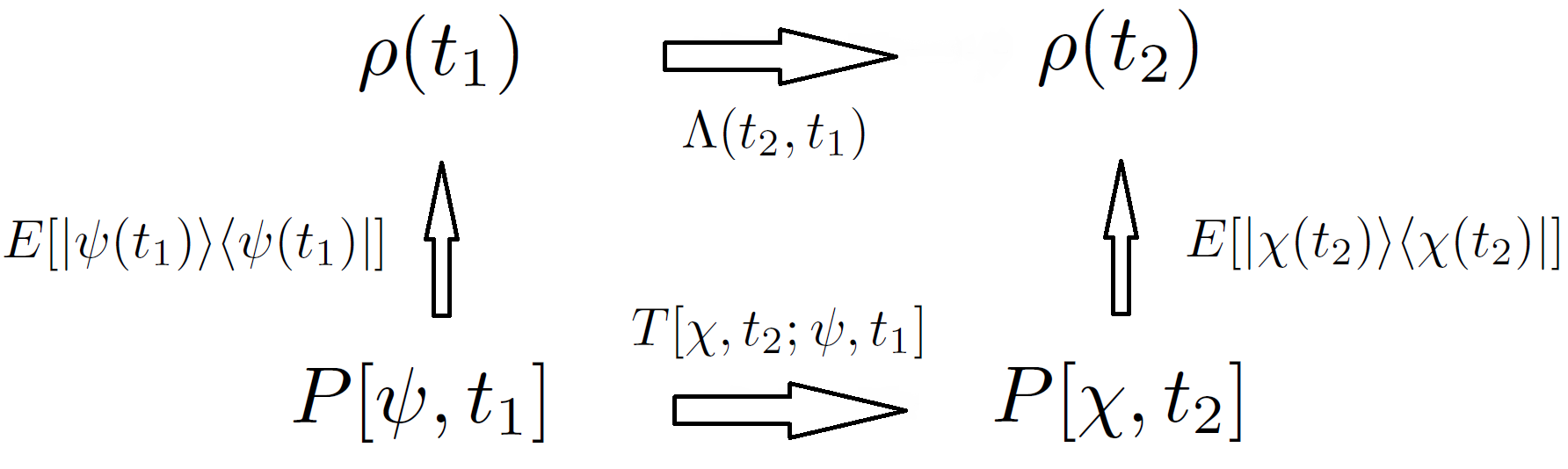}
\caption{Diagram showing the connection between the probability density, the density matrix and their dynamics generated, respectively, by the propagator $T[\chi,t_2;\psi,t_1]$ and by the dynamical map $\Lambda(t_2,t_1)$.}\label{fig61}
\end{figure}
\end{center}

The time evolution of the probability density, from $t_1$ to $t_2$, is generated by a propagator $T[\chi,t_2;\psi,t_1]$ such that $P[\chi,t_2]=\int\mathrm{d}\psi T[\chi,t_2;\psi,t_1]P[\psi,t_1]$. The relation of this formulation to the one in terms of density operators is illustrated in Fig.~\ref{fig61} (see Ref.~\cite{breub}).

Fixing a particular trajectory from time $t_0$ to time $T$ amounts to specifying a number of jumps $N$ and a set of time instants $\{t_k\}$ ($k=1,\dots,N$) such that $t_0<t_1<\dots<t_N<T\equiv t_{N+1}$, at which the wave function jumps along the channels $\{\gamma_{i_k},A_{i_k}\}$.
These single, random and discontinuous events can not be described within the density matrix formalism which gives the ensemble evolution of a collection of independent identical quantum systems or, which is the same, describes the lack of knowledge about the evolving system before a measurement is performed on it. Indeed, if a system evolves from time $t_0$ to time $t$ under the effect of the interaction with an environment, and we do not perform any kind of measurements before $t$, we do not have access to any kind of information about the state of the system and all we can do is to describe its state in terms of generally mixed density matrices. On the other hand, if the evolution of a system is continuously monitored through measurements on its environment, information about single quantum events are collected all along the dynamics (and not just at the final time $t$) and we have at our disposal more information about the state of the open system. This information, which clearly depends on the measuring scheme employed to monitor the environment, is the core of the physical difference between density matrix formalism and the stochastic wave function method, which is nothing but the theoretical description of such a continuous measuring process on the environment.

Results based on this method provide then new insight into the dynamics of a system, and do not trivially just reproduce the knowledge of the density matrix. The choice of a measuring scheme corresponds to the choice of a particular set of pure states into which to decompose the density matrix itself - and such a set does not need to be orthogonal. In both classical and quantum contexts, choosing a pure state decomposition of a mixed state naturally leads to quantify the information content of such a decomposition by employing the so-called Shannon entropy, which is well known to be different from the von Neumann entropy and to depend on the decomposition itself.

In FTs contexts, it is common to define a backward trajectory as the dissipative process generated by a time inversion of the Hamiltonian. This in turn means that any energy exchanges between system and environment get reversed.

Since we decided to extract information about the system by only measuring the environment, one only detects transitions of the open system. Therefore, the backward trajectory is fixed by the requirement that the open system performs transitions at the same time instants as the forward one with rates $\gamma_{i_k}^b$ and jump operators $B_{i_k}=A_{i_k}^{\dag}$. The reason is that the Lindblad operators in the Markovian master equation~(\ref{ME}) and in the weak coupling limit can be divided into two classes $\{A_i^+\}$ and $\{A_i^-\}$ satisfying the conditions $[H_S,A_i^{\pm}]=\pm\epsilon_iA_i^{\pm}$ and $A_i^{+}=(A_i^-)^{\dag}$ \cite{breub} and they thus describe jumps in which an energy quantum $\epsilon_i$ is absorbed ($A_i$, forward trajectory) or emitted ($A_i^{\dag}$, backward trajectory) by the open system from/into the environment (see Section~\ref{examp} for explicit examples). This is the case in many important experimental setups such as, e.g., the many photodetection schemes often employed. On the other hand, the action of the nonunitary operator $U_{\mathrm{eff}}(t_f,t_s)$ on a state during the drift interval $[t_s,t_f]$ reduces its norm in time, describing the decrease of probability of the no-jump event. Therefore, as the backward process itself is a physical dissipative process detected by measurements, its associated drift operator $\mathcal{U}_{\mathrm{eff}}(t_f,t_s)$ has to describe such a decrease of probability along the backward drifts, taking into account that a backward drift propagates the state of the system from time $t_f$ to time $t_s$ such that $t_f>t_s$. Therefore, the Hermitian part of the operator generating backward evolutions has to be unchanged, but its nonhermitian part has to be sign-reversed: This is achieved if one defines $\mathcal{U}_{\mathrm{eff}}(t_f,t_s)=U^{\dag}_{\mathrm{eff}}(t_s,t_f)$. This means, however, that the final state of the backward process may be different from the initial state of the forward one, as in general $A_i^{\dag}A_i\neq\mathbb{I}$ and $\mathcal{U}_{\mathrm{eff}}(t_N,T)U_{\mathrm{eff}}(T,t_N)\neq\mathbb{I}$.

\begin{center}
\begin{figure}[h!]
\includegraphics[width=220pt]{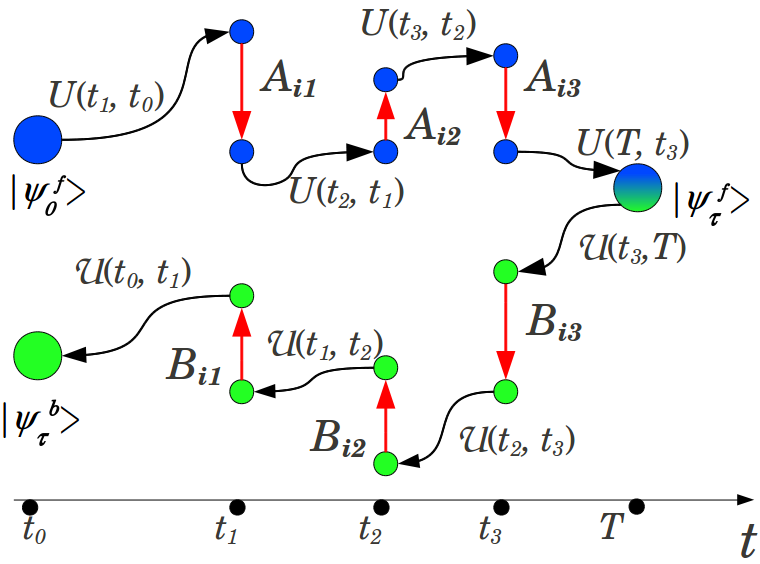}
\caption{(Color online) Pictorial representation of a forward quantum trajectory (dark blue states) and its backward counterpart (light green states) consisting of $N=3$ jumps (vertical red arrows). The backward process is characterized by jumps through the channels $\{\gamma_{i_k}^b,B_{i_k}=A_{i_k}^{\dag}\}$ and by deterministic evolution periods according to $\mathcal{U}_{\mathrm{eff}}(t_k,t_{k+1})=U_{\mathrm{eff}}^{\dag}(t_{k+1},t_k)$.}\label{process}
\end{figure}
\end{center}

In this paper we denote by $|\chi_k^{f(b)}\rangle$ the normalized state of the forward (backward) process right before the jump at time $t_k$, and by $|\psi_k^{f(b)}\rangle$ the normalized state of the forward (backward) process right after the jump at time $t_k$. Exemplary trajectories are schematically depicted in Fig.~\ref{process}, where the forward process starts in the state $|\psi_0^f\rangle$ and ends in $|\psi_{\tau}^f\rangle$, while the final state of the backward one is $|\psi_{\tau}^b\rangle$. By definition $|\psi_{\tau}^f\rangle\equiv |\psi_{0}^b\rangle$ (i.e., the initial state of the backward process) but, in general, $|\psi_0^f\rangle\neq|\psi_{\tau}^b\rangle$ due to the fact that, as already remarked, $A_i^{\dag}A_i\neq\mathbb{I}$ and $\mathcal{U}_{\mathrm{eff}}(t_N,T)U_{\mathrm{eff}}(T,t_N)\neq\mathbb{I}$.
Our goal is to derive an integral FT for entropy production along nonequilibrium processes of this kind. Therefore, we aim at giving explicit expressions for entropy contributions along PDPs.

\section{Entropy}\label{entropy}
A single quantum trajectory (either forward or backward), being a nonequilibrium process, is characterized by a nonzero entropy production.
In the ensemble picture, the entropy of a system is given by its von Neumann entropy $S_{\mathrm{vN}}=-\mathrm{Tr}(\rho\ln\rho)$.
However, as commented in Section~\ref{stochwave}, when employing a quantum unraveling procedure, one describes a single realization of the system dynamics under a particular, fixed measurement scheme. 

Therefore, one has to quantify the amount of information extractable about a particular system when its environment is monitored or, which is the same, the amount of information available in a particular pure state decomposition of a density matrix (such a pure state decomposition corresponds to the set of all possible trajectories generated by the fixed measuring scheme). In quantum contexts, a natural way to quantify the information content of such a decomposition is the so-called Shannon entropy $S=-\int\mathrm{d}\psi P[\psi,t]\ln P[\psi,t]$. Consequently, the single trajectory contribution to the entropy can be defined as $S[\psi]=-\ln P[\psi,t]$. Such an entropy, which clearly depends on the chosen decomposition, has been used in various contexts \cite{mahler, plastina} to quantify information beyond the standard von Neumann one: It is worth stressing that a given decomposition yields more information on the system than the one available in the density matrix only \cite{breub}. In what follows, we will refer to such a quantity as \textit{quantum entropy}. Note that its mathematical definition is formally analogous to the one employed for entropy in classical stochastic processes \cite{seif}.
On the ensemble level the time derivative of the open system quantum entropy is
given by
\begin{equation}\label{dots}
 \dot{S}=-\int\mathrm{d}\psi\dot{P}[\psi,t]\ln P[\psi,t].
\end{equation}
Such a definition is the natural quantum extension of the one employed in many previous works on entropy FTs \cite{esporev, muka, seif}, but it has no classical analogue as it does not reduce to the usual form of entropy in the classical limit. It describes the knowledge about the open quantum system, extracted by an external observer measuring the environment and, as single realizations of quantum dynamics are fundamentally different from their classical counterpart, the information thus extracted can not in general be given any classical interpretation.
Exploiting the explicit form of the master equation for $P[\psi,t]$ \cite{breub}, it is possible to show that the single trajectory contribution to Eq.~(\ref{dots}) can be written as $\dot{S}[\psi]=\dot{S}_j[\psi]+\dot{S}_d[\psi]$, i.e. as the sum of two terms, one arising from the drift part of the PDP (describing the conditioned no-jump evolution of the open system) and one due to the open system jumps.
Since both quantum jumps and drifts are detected by measurements, each of these terms describes a change in knowledge of the external observer about the open system. In addition, we show in what follows that both the jump and the drift entropic terms contribute to entropy production along the nonequilibrium process.

\subsection{Entropy production}
As $\int\dot{S}_j[\psi]\mathrm{d}t$ and $\int\dot{S}_d[\psi]\mathrm{d}t$ only take into account the difference of entropy between initial and final states, but not the features of the transition connecting them, these terms then do not fully describe the information content of unraveling measurements: There are indeed two corresponding terms describing information about transitions, detected in the environment, and which correspond then to entropy flowing from the open quantum system to its bath. Since the full system is out of equilibrium the changes in open quantum system entropy and the flux to the bath are not the same in absolute value. Their difference is interpreted as a net total entropy production along a trajectory, and it is written as
\begin{equation}\label{sigmatot}
\sigma=\Delta S[\psi]-\Delta S^e[\psi],
\end{equation}
$\Delta S^e[\psi]$ being the total entropy flux to the bath and $\Delta S[\psi]=\int_{t_0}^T\mathrm{d}t\dot{S}[\psi]$. In addition, we define a single trajectory jump entropy production and a single trajectory drift entropy production as, respectively,
\begin{eqnarray}
 \sigma_j&=\int_{t_0}^T\mathrm{d}t\dot{S}_j[\psi]-\Delta S^e_j[\psi], \label{sigmaj}\\
 \sigma_d&=\int_{t_0}^T\mathrm{d}t\dot{S}_d[\psi]-\Delta S^e_d[\psi].\label{sigmad}
\end{eqnarray}
In the rest of this work, we aim at giving an expression for the entropy production along a generic quantum trajectory, being thus valid both for what we defined as forward trajectory and for its backward counterpart.

\subsection{Jump entropy production}
Along a generic trajectory, a transition $|\chi_k\rangle\rightarrow|\psi_k\rangle=\frac{A_{i_k}|\chi_k\rangle}{||A_{i_k}|\chi_k\rangle||}$ is characterized by a rate
\begin{equation}
 R_{i_k}^D[\chi_k]=\gamma_{i_k}||A_{i_k}|\chi_k\rangle||^2. \label{rif}
\end{equation}
In what follows, we refer to such a transition as \textit{direct jump}. A direct jump is nothing but the transition experimentally detected within an unraveling approach while following a particular nonequilibrium process (which can, in turn, either be a forward or a backward trajectory). In contrast to a direct jump, we define also a \textit{reversed jump} as $|\psi_k\rangle\rightarrow|\xi_k\rangle=\frac{A_{i_k}^{\dag}|\psi_k\rangle}{||A_{i_k}^{\dag}|\psi_k\rangle||}$, which represents the reversed transition associated to the $k$-th direct jump, and which is a fictitious transition as it is \textit{not} detected in the trajectory: It represents a tool which allows us to introduce a ``direction'' of a single jump and thus its entropy production and, as such, it is intrinsically different from the backward process previously introduced. A reversed jump is associated with the rate
\begin{equation}
 R_{i_k}^R[\chi_k]=\gamma_{i_k}^b\frac{\langle\chi_k|\big(A_{i_k}^{\dag}A_{i_k}\big)^2|\chi_k\rangle}{||A_{i_k}|\chi_k\rangle||^2}\label{rib2}.
\end{equation}
In analogy with classical systems \cite{lebo}, we define the \textit{jump entropy flux} along a single full quantum trajectory as $\Delta S^e_j[\psi]=-\sum_{k=1}^N\ln\frac{R_{i_k}^D[\chi_k]}{R_{i_k}^R[\chi_k]}$.
Note that this definition, despite being formally analogous to the one usually employed in FT contexts when considering pure jump processes \cite{muka, kawa, lnbm}, differs from it because of the structure of transition rates in Eqs.~(\ref{rif}) and~(\ref{rib2}).
The total change of the open system entropy along the process, due to jumps only, is $\Delta S_j[\psi]=-\sum_{k=1}^N\ln \frac{P[\psi_k,t_k]}{P[\chi_k,t_k]}$.
As a consequence, we obtain the total \textit{jump entropy production} along a full quantum trajectory consisting of $N$ jumps as
\begin{equation}\label{sj}
 \sigma_j=\ln\prod_{k=1}^N\frac{R_{i_k}^D[\chi_k]}{R_{i_k}^R[\chi_k]}\frac{P[\chi_k,t_k]}{P[\psi_k,t_k]}.
\end{equation}

\subsection{Drift entropy production}
In order to obtain a time local entropy balance equation for the drift contribution we subdivide each finite drift interval $\big[t_{k-1},t_k\big]$ into many small steps of size $\delta t$. In each of these small time intervals the monitoring of the
environment yields the result that no jump with any of the Lindblad operators
$A_i$ occurs. Conditioned on these events the state vector undergoes small
changes which lead in the limit $\delta t \rightarrow 0$ to a smooth time evolution
describing the drift process.
The formulation is thus analogous to the one given for jump entropy contributions, provided one uses the correct expression for the drift probabilities. The latter are given by $D_{\delta t}^{D(R)}[\psi]=1-\Gamma^{D(R)}[\psi]\delta t$, where $\Gamma^{D(R)}[\psi]=\sum_iR_i^{D(R)}[\psi]$ is the total direct (reversed) jump rate for the state $|\psi\rangle$. The bath entropy contribution of each of these no-jump events is thus $\delta S_d^e[\psi]=-\ln\frac{1-\Gamma^{D}\delta t}{1-\Gamma^{R}\delta t}$.
In this formulation, $\delta t$ is the time interval between two subsequent measurements on the environment. Moreover, since unraveling approaches correspond to continuous measuring processes, it is justified to assume such a time interval to be very small (usually lower bounded only by the resolution time of the measuring apparatus), such that $\Gamma^{D(R)}\delta t\ll 1$.
Under this approximation, we have $\Delta S_d^e[\psi]\sim\int_{t_0}^T\mathrm{d}t\Big(\Gamma^{D}(t)-\Gamma^{R}(t)\Big)$.
Exploiting Eqs.~(\ref{rif}) and~(\ref{rib2}), one can easily prove that 
\begin{eqnarray}
\int_{t_{k-1}}^{t_k}\mathrm{d}t\Gamma^D(t)&=-\ln||U_{\mathrm{eff}}(t_k,t_{k-1})|\psi_{k-1}\rangle||^2,\\
\int_{t_{k-1}}^{t_k}\mathrm{d}t\Gamma^R(t)&=-\ln||\mathcal{U}_{\mathrm{eff}}(t_{k-1},t_k)|\psi_{k}\rangle||^2,
\end{eqnarray}
so that $\Delta S^e_d[\psi]=-\ln\prod_{k=1}^{N+1}\frac{||U_{\mathrm{eff}}(t_k,t_{k-1})|\psi_{k-1}\rangle||^2}{||\mathcal{U}_{\mathrm{eff}}(t_{k-1},t_k)|\psi_{k}\rangle||^2}$.
The total drift-induced change of open quantum system entropy is $\Delta S_d[\psi]=-\sum_{k=1}^{N+1}\ln \frac{P[\chi_k,t_k]}{P[\psi_{k-1},t_{k-1}]}$ and finally
\begin{equation}\label{sd}
 \sigma_d=\ln\prod_{k=1}^{N+1}\frac{||U_{\mathrm{eff}}(t_k,t_{k-1})|\psi_{k-1}\rangle||^2}{||\mathcal{U}_{\mathrm{eff}}(t_{k-1},t_k)|\psi_{k}\rangle||^2}\frac{P[\psi_{k-1},t_{k-1}]}{P[\chi_k,t_k]}
\end{equation}
is the single trajectory drift entropy production.
With the use of Eqs.~(\ref{sigmatot}),~(\ref{sj}) and~(\ref{sd}) it is straightforward to show that
\begin{equation}\begin{split}\label{sigma}
\sigma&\equiv \sigma_j + \sigma_d =\ln\bigg(\frac{P[\psi_0,t_0]}{P[\chi_{N+1},t_{N+1}]}\\
&\times\prod_{k=1}^N\frac{R^D_{i_k}[\chi_k]}{R^R_{i_k}[\chi_k]}\prod_{k=1}^{N+1}\frac{||U_{\mathrm{eff}}(t_k,t_{k-1})|\psi_{k-1}\rangle||^2}{||\mathcal{U}_{\mathrm{eff}}(t_{k-1},t_k)|\psi_{k}\rangle||^2}\bigg).
\end{split}\end{equation}
Such an equation describes the total entropy production along a single quantum trajectory: In particular, since a quantum trajectory is followed by measuring the environment, $\sigma$ is the total information the external observer acquires about the system through the knowledge of initial and final states of the process ($\Delta S$) \textit{minus} the information extracted by measurements of all intermediate steps connecting them ($\Delta S^e$), detected in the bath. Note that Eq.~(\ref{sigma}) is fully characterized by the knowledge of a single trajectory, contrarily to the single trajectory contribution to von Neumann entropy which requires the solution of the full master equation of the system.

\subsection{Entropy flux and quantum fluctuations}\label{quantumfluct}
To fully understand the physics described by the entropy flux terms introduced in above, let us analyze for instance its jump contribution in Eq.~(\ref{sj}). In the case of a jump $|\chi_k\rangle\rightarrow|\psi_k\rangle$ the entropy flowing into the environment is given by
\begin{equation}\label{deltasejk}
\Delta S_{j_k}^e=\ln\frac{\gamma_{i_k}^b}{\gamma_{i_k}}+\ln\frac{\gamma_{i_k} R^R_k}{\gamma_{i_k}^b R^D_k}.
\end{equation}
On average the process has a preferred direction if the two rates are not equal. Since, in a weak coupling Markovian master equation with a thermal environment, $\gamma_{i_k}^b/\gamma_{i_k}=e^{-\beta \epsilon_{i_k}}$ ($\epsilon_{i_k}$ being the energy $Q_E$ exchanged between system and environment during the transition $A_{i_k}$), the first term on the r.h.s. of Eq.~(\ref{deltasejk}) is a standard thermodynamic entropic flux of the form $-\frac{Q_E}{T}$. The second term on the r.h.s. of Eq.~(\ref{deltasejk}) describes, on the other hand, how much information is produced by the system jumping through the \textit{particular} decay channel $A_{i_k}$. We refer to such an additional term as \textit{nonthermal entropy flux} $\Delta S^{nt}$. We can characterize such a nonthermal flux by introducing the parameter $\eta_k=1-\gamma_{i_k} R^R_k/\gamma_{i_k}^b R^D_k$.
According to its definition, $\eta_k=0$ if the bias of the associated direct transition to the corresponding reversed one is only due to the direction of heat flux. Introducing the operator $\Lambda_{i_k}=A_{i_k}^{\dag}A_{i_k}$ and exploiting the explicit expression of $R^R$ and $R^D$ one obtains
\begin{equation}\label{etak}
 \eta_k=\frac{\langle\chi_k|\Lambda_{i_k}|\chi_k\rangle^2-\langle\chi_k|\Lambda_{i_k}^2|\chi_k\rangle}{||A_{i_k}|\chi_k\rangle||^4}=-\frac{\mathrm{Var}_1^{[\chi_k]}(\Lambda_{i_k})}{||A_{i_k}|\chi_k\rangle||^4},
\end{equation}
where $\mathrm{Var}_1(Q)=\int\mathrm{d}\psi P[\psi]\Big(\langle\psi|Q^2|\psi\rangle-\langle\psi|Q|\psi\rangle^2\Big)$, introduced in \cite{hpvar}, is known to measure the average intrinsic quantum fluctuations of an operator $Q$ during a dynamic process, and $\mathrm{Var}_1^{[\chi_k]}(Q)=\langle\chi_k|Q^2|\chi_k\rangle-\langle\chi_k|Q|\chi_k\rangle^2$ is its single trajectory contribution due to the $k$-th jump. From the structure of $\eta_k$ we infer that during the jump $|\chi_k\rangle\rightarrow|\psi_k\rangle$, the exchange of information between system and environment goes beyond the standard thermodynamic form if and only if the operator $\Lambda_{i_k}$ has nonzero purely quantum fluctuations in the source state of the direct jump: The nonthermal entropic contribution has indeed the form $\Delta S_{j_k}^{nt}=\ln(1-\eta_k)$.

The additional, nonthermal contribution to the jump entropy flux is directly linked to the quantum fluctuation of the operators $\Lambda_{i_k}$, which shows the nonclassical character of our results.
Note that, thanks to the same formal structure of the jumps and the drifts transition rates, these results hold true also for the drift parts of a quantum trajectory. In particular, during a drift there is no standard thermodynamic entropy flux as the heat flux vanishes. However, thanks to the purely quantum fluctuations of the operator $\Omega_k=U_{\mathrm{eff}}^{\dag}(t_k,t_{k-1})U_{\mathrm{eff}}(t_k,t_{k-1})$ in the state $|\psi_{k-1}\rangle$, the generic $k$-th drift part of the full process is also associated to a purely quantum information flux between system and environment. It is worth stressing however that the non-thermal drift entropy flux is of the order of $\delta t^2$ ($\delta t$ being the time interval between two subsequent measurements), while the corresponding jump term does not depend on $\delta t$. The non-thermal drift term is analysed in more detail in the
Appendix.

\section{Integral Fluctuation Theorem}\label{IFT}
We investigate the statistical properties of $\sigma\equiv\sigma_f$ in Eq.~(\ref{sigma}) along a forward process.
To simplify the notation, in what follows we introduce the symbols $D_k^{D}[\psi_k]=||U_{\mathrm{eff}}(t_k,t_{k-1})|\psi_{k-1}\rangle||^2$ and $D_k^{R}[\psi_k]=||\mathcal{U}_{\mathrm{eff}}(t_{k-1},t_{k})|\psi_{k}\rangle||^2$. Moreover, rates along forward or backward trajectories will be denoted by specifying the trajectory directly in the functional dependence of the rates on the wave function, so that for example $R_k^{R(D)}[\chi_k^{f(b)}]$ is the reversed (direct) $k$-th jump rate of the forward (backward) trajectory. With these notations, the mean value of $e^{-\sigma_f}$ (commonly considered in FTs contexts) can be evaluated as
\begin{equation}\begin{split}
 \big\langle e^{-\sigma_f}\big\rangle&\equiv\int\mathrm{d}\psi^f P[\psi^f]e^{-\sigma[\psi^f]}\delta\big(\sigma[\psi^f]-\sigma_f\big)\\
&=\int\mathrm{d}\psi^b P[\psi^b] \prod_{k=1}^N\frac{R^R_{i_k}[\chi_k^f]}{R^D_{i_k}[\chi_k^b]}\frac{D_k^R[\psi_k^f]}{D_k^D[\psi_k^b]}
\end{split}\end{equation}
and, since $P[\psi^b]$ is by construction a normalised probability distribution, one obtains
\begin{equation}\label{totft}
 \big\langle e^{-\sigma_f}\big\rangle=\Bigg\langle\prod_{k=1}^N\frac{R^R_{i_k}[\chi_k^f]}{R^D_{i_k}[\chi_k^b]}\frac{D_k^R[\psi_k^f]}{D_k^D[\psi_k^b]}\Bigg\rangle=1+\zeta_f,
\end{equation}
where $\langle\cdot\rangle$ stands for an average over all possible realizations of a nonequilibrium process.
Equation (\ref{totft}) shows that, in the case of quantum trajectories, $\langle e^{-\sigma_f}\rangle$ is not a universal constant: The r.h.s. is indeed, in general, different from 1 and depends on the set of Lindblad operators characterizing the unraveling scheme. This results in a quantum correction $\zeta_f$ to the classical result. We expect such a correction to be positive: Since, as commented previously, $\sigma_f$ is the difference between the information extracted only by measuring initial and final states of a trajectory and the information available by following the full quantum trajectory, it is reasonable to expect that the latter is greater than the former. Therefore, on average we have $\langle\sigma_f\rangle<0$ which leads to $\zeta_f>0$. This is illustrated in Section~\ref{examp}, where we study the predictions of Eq.~(\ref{totft}) numerically for several model systems. In particular, such a correction originates from the fundamental difference between a backward process (which is a real dissipative process) and ``reversed'' processes (which is the collection of all reversed processes and, as such, is fictitious).

Such a distance is nothing but the consequence of the measuring scheme employed to characterize trajectories: Information acquired about the system by the external observer is not symmetric under time reversal, and such a broken symmetry of knowledge produces different states in forward and backward transitions. This physically results in the presence of the nonthermal quantum entropic flux~(\ref{deltasejk}), which does not obey a standard FT. Indeed it has recently been shown \cite{pek} that, if only thermal energy exchanges during jumps are taken into account along quantum trajectories of an open two-level system, the standard universal form of FT holds. In addition, a recently published work \cite{horow} showed that the choice of a particular measuring scheme can lead to a standard entropic FT.
As a matter of fact, in a ``standard''-like limit the nonthermal entropy flux vanishes both for drifts and jumps due to the fact that the operators $\Omega_k$ and $\Lambda_{i_k}$ have vanishing quantum fluctuations, and in this case $\zeta_f=0$ recovering the universal standard form of FT.

\section{Examples}\label{examp}
In this Section we present some results on particular systems exemplifying our findings. The first example shows a particular limit case in which the quantum correction $\zeta_f$ vanishes. In the second example we discuss how the choice of a particular measuring scheme affects entropic quantum fluctuations resulting in deviations from the standard FT.

\subsection{The standard case: jumps between free Hamiltonian eigenstates}
As an example of the standard limit of our results we mentioned in Section~\ref{IFT} an open system without driving, whose Lindblad operators and decay rates remain constant in time. In the Markovian and weak coupling limit, its Lindblad operators satisfy $[H_S,A_k^{\pm}]=\pm\epsilon_kA_k^{\pm}$. If now we assume the free Hamiltonian $H_S$ to have nondegenerate energy gaps in its spectrum (this assumption is typically employed when studying, e.g., quantum thermalization processes \cite{therm}), the emission of an energy quantum $\epsilon_i$ is in a one-to-one correspondence with a transition between two well defined energy levels $|n\rangle$ and $|m\rangle$ such that $\omega_{n}-\omega_m=\epsilon_k$, $\omega_i$ being the energy associated to the eigenstate $|i\rangle$ of $H_S$. Assuming the spectrum of $H_S$ is composed of $N$ discrete levels ($|1\rangle$ being the ground state) of increasing energy, the natural choice for the set of Lindblad operators is then
\begin{eqnarray}
 A_{N(i-1)+j-\frac{i(i+1)}{2}}&=|i\rangle\langle j|\,\,\,\mathrm{for}\,\,1\leq i<j\leq N,\label{setA}\\
 A_{N(i-1)+j-\frac{i(i+1)}{2}}^{\dag}&=|j\rangle\langle i|\,\,\,\mathrm{for}\,\,1\leq i<j\leq N.\label{setAd}
\end{eqnarray}
Note that, thanks to the assumption of nondegenerate gaps in $H_S$ and the form of the operators in the set $\{A_k\}$, it is not necessary for the system to start its trajectory in an eigenstate of $H_S$ since after the first jump any wave function $|\psi\rangle$ is projected to a well defined energy eigenstate. We can therefore assume, without loss of generality, that the system starts its trajectory from a generic yet fixed energy eigenstate $|n\rangle$. The action of a jump operator $|m\rangle\langle n|$ on such a state is then nothing but the transition $|n\rangle\rightarrow|m\rangle$. The system performs then jumps only between eigenstates of its free Hamiltonian. Exploiting Eqs.~(\ref{setA}) and~(\ref{setAd}), one notices that $A_{N(i-1)+j-\frac{i(i+1)}{2}}^{\dag}A_{N(i-1)+j-\frac{i(i+1)}{2}}=|j\rangle\langle j|$, so that the drift non-Hermitian Hamiltonian becomes $H_{\mathrm{eff}}=H_S-\frac{i}{2}\sum_i^N\widetilde{\gamma}_i|i\rangle\langle i|$, where $\widetilde{\gamma}_i=\sum_j^N\gamma_{N(i-1)+j-\frac{i(i+1)}{2}}$ is the total relaxation rate associated with the energy level $|i\rangle$. The drift operator $U_{\mathrm{eff}}(t_{k},t_{k-1})$ is then diagonal in the eigenbasis of $H_S$ and introduces nothing but a phase factor to any evolving energy eigenstates. Any trajectory of this kind is equivalent to a pure jump process between energy eigenstates. We note two things: On the one hand, since the emission or absorption of an energy quantum always connects the same two states, and since drifts have no effects on the trajectory, backward and reversed processes are the same and the backward trajectory connects the same states as the forward one, but in reversed order. This in turn means that the quantum correction $\zeta_f$ in Eq.~(\ref{totft}) vanishes, and one recovers the standard form of fluctuation theorems. On the other hand, as expected, this is due to the fact that nonthermal entropic fluxes are zero, since it can be straightforwardly shown that neither the operators $\Lambda_{N(i-1)+j-\frac{i(i+1)}{2}}=|j\rangle\langle j|$ nor the operators $U^{\dag}_{\mathrm{eff}}(t_k,t_{k-1})U_{\mathrm{eff}}(t_k,t_{k-1})=\sum_i^Ne^{-\widetilde{\gamma}_i(t_k-t_{k-1})}|i\rangle\langle i|$ have purely quantum fluctuations in any energy eigenstate, as they are diagonal in such a basis. The process is thus, in this respect, fully classical.

\subsection{Driven two-level atom}\label{supp}
As a more interesting example of our results, we unravel the dynamics of a driven two-level atom ($|e\rangle$ and $|g\rangle$ being, respectively, its excited and ground state) under two different unraveling schemes somehow analogous to, respectively, the one describing a direct photodetection of emitted light and the so-called homodyne photodetection. We assume that the atom interacts with a reservoir of field modes at nonzero temperature. The atomic master equation, 
written in the form of Eq.~(\ref{ME}), is given by
\begin{equation}\begin{split}\label{twolevelME}
 \dot{\rho}(t)&=-i\frac{\omega(t)}{2}[\sigma_x,\rho(t)]\\
 &+\gamma_1(t)\Big(\sigma_-\rho(t)\sigma_+-\frac{1}{2}\big\{\sigma_+\sigma_-,\rho(t)\big\}\Big)\\
 &+\gamma_2(t)\Big(\sigma_+\rho(t)\sigma_--\frac{1}{2}\big\{\sigma_-\sigma_+,\rho(t)\big\}\Big),
\end{split}\end{equation}
where $\omega(t)$ accounts for the applied external driving, $\sigma_-=\sigma_+^{\dag}=|g\rangle\langle e|$ is the lowering operator of the atom and the rates $\gamma_1(t)$ and $\gamma_2(t)$ depend on the atom-field coupling parameter, on the structure of the state of the field and on its spectrum. Note that, as long as $\gamma_1(t),\gamma_2(t)\geq 0\,\,\,\,\forall t$, Eq.~(\ref{twolevelME}) always implements a time-dependent Markovian dynamics.
The  ``direct photodetection-like`` unraveling yields two jump operators of the form
\begin{eqnarray}
 A_1&=&\sigma_-,\label{directa1}\\
 A_2&=&\sigma_+=A_1^{\dag},\label{directa2}
\end{eqnarray}
describing respectively emission and absorption of a quantum of light by the atom, with relaxation rates $\gamma_1(t)$ (emission) and $\gamma_2(t)$ (absorption).

On the other hand another suitable set of Lindblad operators, similar to the ones describing the "homodyne" photodetection process, is given by
\begin{eqnarray}
 A_1^-(\beta)&=&\sigma_--i \beta,\label{homodynea1}\\
 A_1^+(\beta)&=&\sigma_-+i \beta,\label{homodynea2}\\
 A_2^-(\beta)&=&\sigma_+-i \beta^*=A_1^+(\beta)^{\dag},\label{homodynea3}\\
 A_2^+(\beta)&=&\sigma_++i \beta^*=A_1^-(\beta)^{\dag},\label{homodynea4}
\end{eqnarray}
for any $\beta \in \mathbb{C}$. The associated relaxation rates are $\gamma_1^{\pm}(t)=\frac{\gamma_1(t)}{2}$ and $\gamma_2^{\pm}(t)=\frac{\gamma_2(t)}{2}$. Note that the transformation of Lindblad operators leading to the set~(\ref{homodynea1})-(\ref{homodynea4}) produces no changes in the Hamiltonian part thanks to the fact that $A_1^-(\beta)+A_1^+(\beta)=2 A_1$ and $A_2^-(\beta)+A_2^+(\beta)=2 A_2$. It is easy to check that the master equation obtained using the four operators~(\ref{homodynea1})-(\ref{homodynea4}) reduces, for any $\beta$, to Eq.~(\ref{twolevelME}), therefore describing the same physical process on the ensemble level. Fixing $\beta$ one fixes a particular measuring scheme and, therefore, a particular set of Lindblad operators. In this way we are able, just by switching between the two sets~(\ref{directa1}),~(\ref{directa2}) and~(\ref{homodynea1})-(\ref{homodynea4}) and/or by tuning $\beta$, to investigate the dependence of $\zeta_f$ in Eq.~(\ref{totft}) on the unraveling scheme employed.

We have performed simulations for the ``direct photodetection-like`` scheme, and for the ``homodyne-like`` scheme with different values of $\beta$, with fixed measurement step $\delta t$ and total duration $T$, choosing $\omega(t)=\omega_0 \Big(1-e^{-\frac{t}{\tau}}\Big)$, $\gamma_1(t)=g_1 e^{-\frac{t}{\tau_1}}$ and $\gamma_2(t)=g_2\Big(1-e^{-\frac{t}{\tau_2}}\Big)$. The parameters have been fixed such that $\frac{\delta t}{\tau_1}=1.3*10^{-3}, \frac{\delta t}{\tau_2}=10^{-3}, \frac{\delta t}{\tau}=2.7*10^{-3}$ and $\frac{\delta t}{T}=8*10^{-4}$. The initial atomic wave function is of the form $|\psi_0^f\rangle=c_e|e\rangle+c_g|g\rangle$ and, for each trajectory, the complex values for $c_e$ and $c_g$ have been chosen randomly out of a uniform distribution of real values in $[0,1]$ for their moduli, and of a uniform distribution of real angles in $[0,2\pi]$ for their relative phase. Note that such a distribution does not correspond to a uniform distribution of pure states over the Bloch sphere. We stress that, at least in principle, any distribution of state vectors can be generated by appropriate preparation measurements.

\begin{center}
\begin{figure}[h!]
\includegraphics[width=250pt]{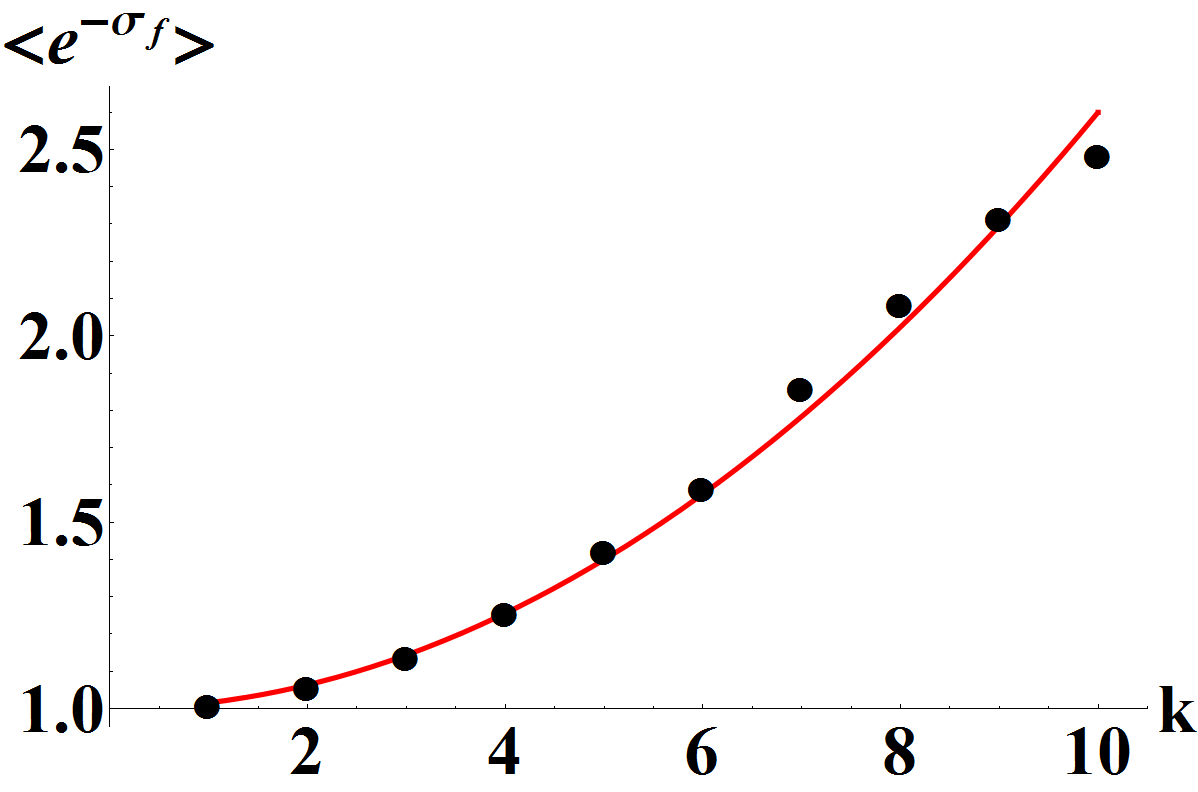}
\caption{(Color online) $\big\langle e^{-\sigma_f}\big\rangle$ (black dots), evaluated over $10^4$ quantum trajectories for the ``direct photodetection-like`` scheme for a driven two-level atom interacting with a reservoir of modes, for 10 different values of driving amplitude and relaxation rates $\delta t \omega_0=8k*10^{-4}, \,\delta t g_1=4.8k*10^{-4},\, \delta t g_2=8k*10^{-5}$ for $k=1,\dots,10$ and having fixed the other parameters as $\frac{\delta t}{\tau_1}=1.3*10^{-3}, \frac{\delta t}{\tau_2}=10^{-3}, \frac{\delta t}{\tau}=2.7*10^{-3}$ and $\frac{\delta t}{T}=8*10^{-4}$. The red full line is a quadratic function of $k$ roughly interpolating numerical data and their increasing trend.}\label{directft}
\end{figure}
\end{center}

\begin{center}
\begin{figure}[h!]
\includegraphics[width=250pt]{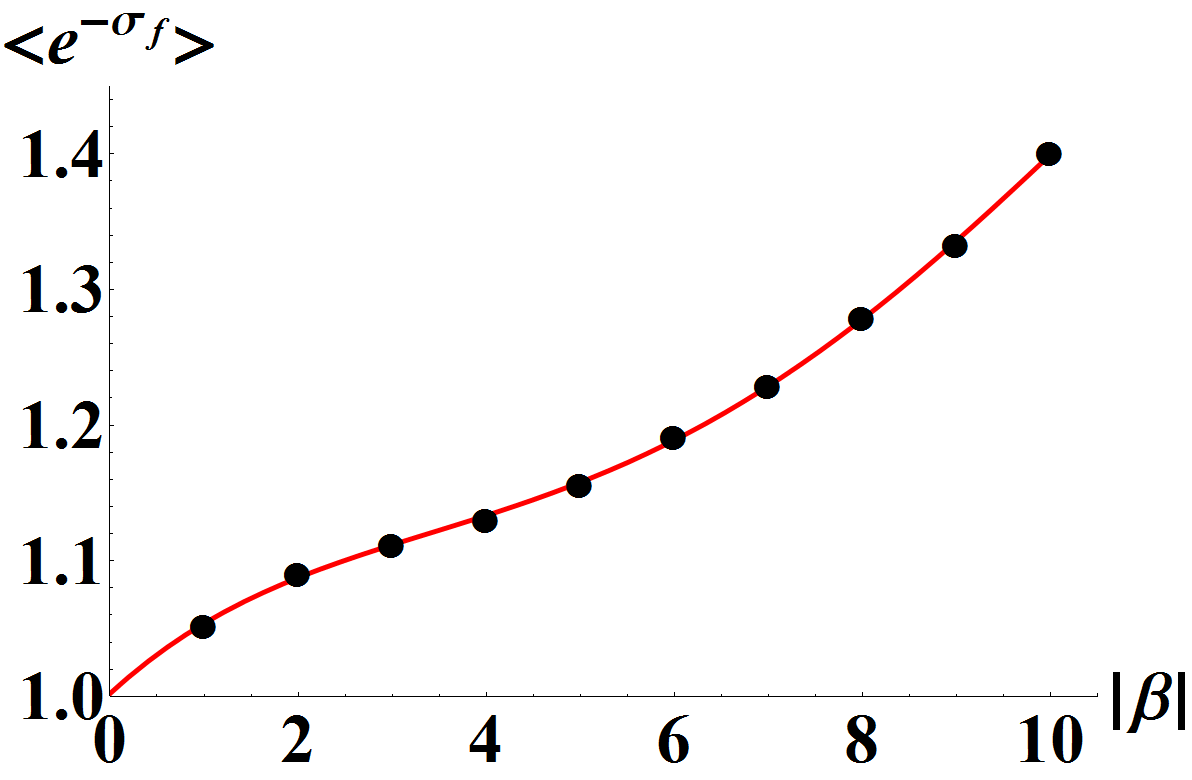}
\caption{(Color online) $\big\langle e^{-\sigma_f}\big\rangle$ (black dots), evaluated over $10^4$ quantum trajectories for the "homodyne-like" scheme for a driven two-level atom interacting with a reservoir of modes, for 10 different values of $\beta=k e^{\frac{i 3\pi}{5}},\,\,k=1,\dots,10$ and with fixed system parameters as $\delta t\omega_0=5.6*10^{-4}, \delta tg_1=4*10^{-4}, \delta tg_2=2.4*10^{-4}, \frac{\delta t}{\tau_1}=1.3*10^{-3}, \frac{\delta t}{\tau_2}=10^{-3}$ and $\frac{\delta t}{\tau}=2.7*10^{-3}$. The red full line is a $4$th degree polynomial function of $|\beta|$ roughly interpolating numerical data and their increasing trend.}\label{homodyneft}
\end{figure}
\end{center}

The results of these simulations are shown in Figs.~\ref{directft} and~\ref{homodyneft} where $\big\langle e^{-\sigma_f}\big\rangle$, evaluated as an average over $10^4$ quantum trajectories, is shown for, respectively, 10 different sets of values of rates and driving such that $\delta t \omega_0=8k*10^{-4}, \,\delta t g_1= 8k*10^{-5},\, \delta t g_2=4.8k*10^{-4}$ for $k=1,\dots,10$ (``direct photodetection-like`` scheme, Fig.~\ref{directft}) or 10 different values of $\beta=k e^{\frac{i 3\pi}{5}},\,\,k=1,\dots,10$ and $\delta t\omega_0=5.6*10^{-4}, \delta tg_1=4*10^{-4}, \delta tg_2=2.4*10^{-4}$ (''homodyne-like'' scheme, Fig.~\ref{homodyneft}).

\begin{center}
\begin{figure}[h!]
\includegraphics[width=250pt]{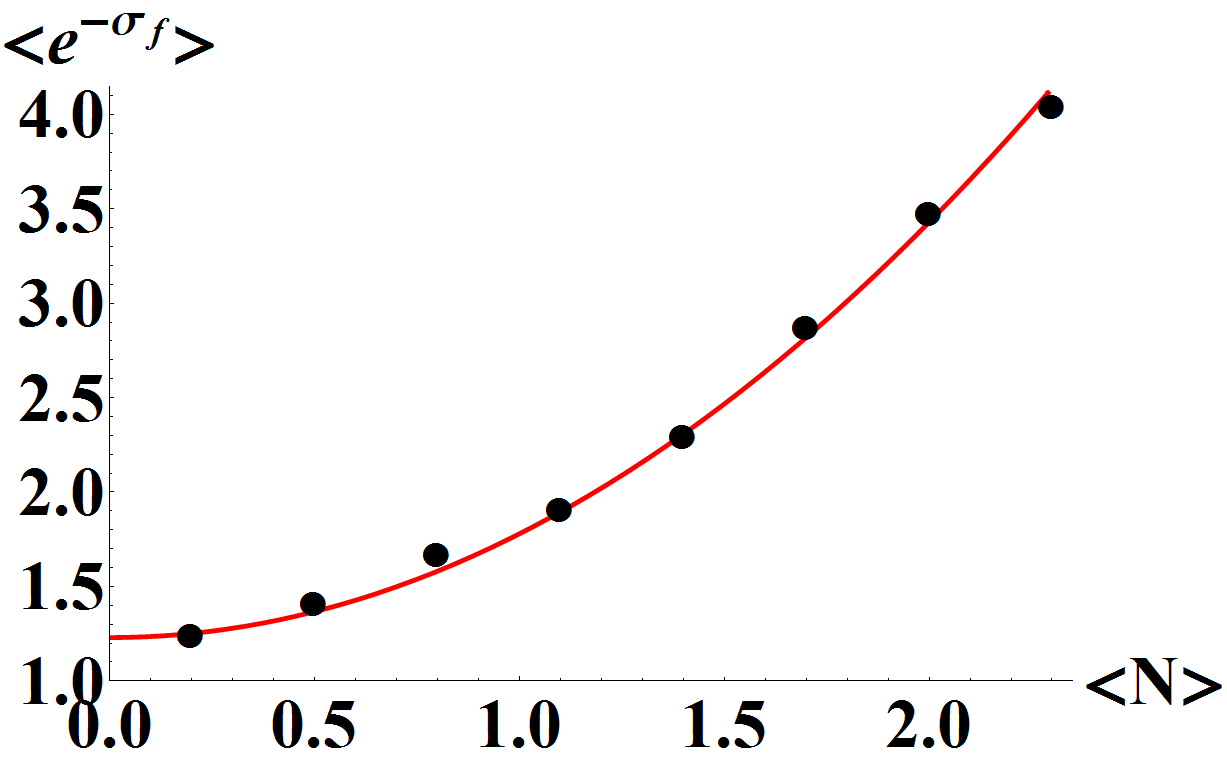}
\caption{(Color online) $\big\langle e^{-\sigma_f}\big\rangle$ (black dots), evaluated over $3*10^4$ quantum trajectories for the ``direct photodetection-like`` scheme for a driven two-level atom interacting with a thermal reservoir of modes, for 8 different values of temperature such that the average thermal photon number is $\langle N\rangle=0.2+0.3k,\,\,k=0,\dots,7$ and having fixed the other parameters as $\frac{\delta t}{\tau}=2.7*10^{-3},\omega_0 \delta t=8*10^{-4}$ and $\frac{\delta t}{T}=8*10^{-4}$. The red full line is a quadratic function of $\langle N\rangle$ roughly interpolating numerical data and their increasing trend.}\label{directft2}
\end{figure}
\end{center}

Finally, we analyze the more familiar case in which the values of decay rates are determined by environmental properties only, i.e. the case of a thermal bath weakly interacting with the system: Fig.~\ref{directft2} shows results for time independent relaxation rates $\gamma_1\propto \langle N\rangle+1$ and $\gamma_2 \propto \langle N\rangle$, $\langle N\rangle$ being the average photon number in the field state. Note that in this case the explicit functional dependence of $\gamma_1$ and $\gamma_2$ on the properties of a thermal bath (such as, for example, its spectrum or its temperature) can be obtained through the theory of Einstein's coefficients. Indeed the dependence of $\gamma_1$ and $\gamma_2$ on $\langle N\rangle$ describes the effects of atomic absorption and of both spontaneous and stimulated atomic emissions \cite{carmichael}.

This further run of simulations, consisting of $3*10^4$ trajectories
for each point in the plot, has been performed analogously to the
one reported in Fig.~\ref{directft}, keeping all the parameters
fixed at the same value characterizing Fig.~\ref{directft}, with the
only exception of $\omega_0$ which has been fixed such that
$\omega_0 \delta t=8*10^{-4}$ and, of course, the rates $\gamma_1$
and $\gamma_2$. The simulations have been performed for 8 different
values of $\langle N\rangle$ such that $\langle
N\rangle=0.2+0.3k,\,\,k=0,\dots,7$. Note that constant relaxation
rates obeying $\gamma_1=\gamma_2+1$ are obtained for a reservoir of
modes in thermal equilibrium at fixed temperature, in which case
$\gamma_2$ is proportional to the average number of photons in the
thermal state of the field. Tuning $\langle N\rangle$ in our
simulations, therefore, corresponds to tuning the temperature of the
field with which the two-level atom interacts (provided its spectrum
stays constant).

Two interesting features emerge from these simulations: first of all, the mean value $\big\langle e^{-\sigma_f}\big\rangle$ can be substantially different from 1 both for ``direct photodetection-like`` and ``homodyne-like'' schemes, resulting in a nonzero quantum correction $\zeta_f$. Therefore, even for such a simple system the difference between backward trajectory and reversed processes becomes nonneglegible. Secondly, $\big\langle e^{-\sigma_f}\big\rangle$ shows a clear dependence on the set $\{\omega_0, g_1, g_2\}$, on the average bath photon number $\langle N\rangle$ and on $|\beta|$, i.e. on the driving and the strength of the decay, on the bath temperature and on the unraveling scheme employed. In particular, in the ``direct photodetection-like`` scheme $\big\langle e^{-\sigma_f}\big\rangle$ is very close to $1$ in the case of a weakly decaying and driven system ($k=1$) and increases smoothly with $k$ with a power-law like shape. Also in the case of the homodyne-like scheme a clear increasing trend is detected which suggests a monotonic increase of $\big\langle e^{-\sigma_f}\big\rangle$ with $|\beta|$, properly described by a quadratic function of $|\beta|^2$. Finally, it is interesting to note that, in the case of a thermal bath, $\big\langle e^{-\sigma_f}\big\rangle$ increases quadratically with the average photon number $\langle N\rangle$ but does not tend to $1$ for $\langle N\rangle\rightarrow 0$, since also in the case of a zero temperature bath the system can perform quantum jumps and undergoes nontrivial drifts, resulting in a nonvanishing nonthermal entropy flux.
These features may reasonably be employed to properly engineer a class of nonequilibrium processes with particular stochastic properties of entropy production.

\section{Conclusions}\label{conclu}
We have obtained an expression for stochastic entropy production along a purely quantum trajectory of a driven open system, defined through continuous measurements on the environment only. The quantum entropy thus defined, which is fundamentally different from the commonly employed von Neumann entropy, describes the observer's gain/loss of information about the open system along single realizations of quantum dynamics and, contrarily to previous approaches to quantum FTs, does not require any knowledge on the ensemble dynamics of the open system, given by the solution of the master equation~(\ref{ME}). We showed that the flux of such an entropy is not only associated to energy flux from/into the bath, defying common classical thermodynamic expectations. The additional information term results from purely quantum fluctuations of the transition operators along a trajectory. Due to this additional term, the quantum entropy of a stochastic trajectory does not obey the usual form of integral fluctuation theorem: The quantum correction $\zeta_f$ in Eq.~(\ref{totft}) depends on the set of jump operators employed to unravel the master equation, and ultimately describes the difference between the physical backward trajectory and the fictitious reversed processes. In other words, such a correction is due to the lack of symmetry between forward and backward processes, which in turn originates from the existence of an external observer performing measurements on the bath to detect transitions.

\appendix*

\section{Nonthermal drift entropy flux}

In the main manuscript we showed that nonthermal contributions to entropy flux originate from purely quantum fluctuations of a certain kind of operators, either jump operators $A^{\dag}A$ or drift operators $U^{\dag}U$. Here we want to analyze the structure of these fluctuations along nonunitary evolutions. During a drift interval $[t_{k-1},t_k]$, information is extracted by measuring the environment at a constant rate $\frac{1}{\delta t}$. Therefore the total information extracted is given by a sum of small contributions, each of which originates from one of the $\frac{t_k-t_{k-1}}{\delta t}$ measuring processes and originates from the quantum fluctuations of the operator $U^{\dag}(\delta t)U(\delta t)$. We assume here the most general situation, in which $H_{\rm eff}(t)$ depends on time and $\big[H_{\rm eff}(t_1),H_{\rm eff}(t_2)\big]\neq 0$ so that the use of the time-ordering operator $\mathcal{T}$ is needed.
Since in the stochastic wave function method one always assumes $\delta t$ to be very small ($\sum_i R_i\delta t\ll1$), one can keep only terms up to order $\delta t^2$ in the Dyson expansion of the drift operator. For the sake of simplicity, we introduce the Hermitian operator $\Omega(t)=\sum_i\gamma_i(t)A_i^{\dag}(t)A_i(t)$, so that $H_{\rm eff}(t)=H_S(t)-\frac{i}{2}\Omega(t)$, and the operators 
\begin{eqnarray}
I_{\delta t}^{(1)}(t)&=&\int_t^{t+\delta t}\Omega(t_1)\mathrm{d}t_1,\\
I_{\delta t^2}^{(2)}(t)&=&\int\int_t^{t+\delta t}\mathrm{d}t_1\mathrm{d}t_2\mathcal{T}\big\{\Omega(t_1)\Omega(t_2)\big\},\\
T_{\delta t^2}(H_1,H_2)&=&\int\int_t^{t+\delta t}\mathrm{d}t_1\mathrm{d}t_2\Big(\mathcal{T}\big\{H_S(t_1)H_S(t_2)\big\}\nonumber \\
&-&H_S(t_1)H_S(t_2)\Big),\\
V_{\delta t^2}(\Omega_1,\Omega_2)&=&\int\int_t^{t+\delta t}\mathrm{d}t_1\mathrm{d}t_2\Big(\mathcal{T}\big\{\Omega(t_1)\Omega(t_2)\big\}\nonumber \\
&+&\Omega(t_1)\Omega(t_2)\Big),\\
J_{\delta t^2}(\Omega,H)&=&\int\int_t^{t+\delta t}\mathrm{d}t_1\mathrm{d}t_2\Big(\Omega(t_1)H_S(t_2)\nonumber \\
&-&H_S(t_1)\Omega(t_2)\Big).\\
&\nonumber
\end{eqnarray}

With these notations one obtains
\begin{equation}\begin{split}
 U(\delta t)&=\mathcal{T}\exp\left\{-i\int_{t}^{t+\delta t}H_{\rm eff}(t_1)\mathrm{d}t_1\right\}\\
 &\sim 1-\frac{1}{2}\int\int_t^{t+\delta t}\mathrm{d}t_1\mathrm{d}t_2\mathcal{T}\Big\{H_{\rm eff}(t_1)H_{\rm eff}(t_2)\Big\}\\
 &-i\int_t^{t+\delta t}H_{eff}(t_1)\mathrm{d}t_1
\end{split}\end{equation}
and, up to order $\delta t^2$,
\begin{eqnarray}
 U(\delta t)^{\dag}U(\delta t)&\sim& 1-I_{\delta t}^{(1)}(t)-T_{\delta t^2}(H_1,H_2)\nonumber \\
 &+&\frac{1}{4}V_{\delta t^2}(\Omega_1,\Omega_2)+\frac{i}{2}J_{\delta t^2}(\Omega,H),
\end{eqnarray}
and
\begin{eqnarray}
 \Big(U(\delta t)^{\dag}U(\delta t)\Big)^2&\sim 1-2I_{\delta t}^{(1)}(t)+I_{\delta t^2}^{(2)}(t)-2T_{\delta t^2}(H_1,H_2)\nonumber\\
 &+\frac{1}{2}V_{\delta t^2}(\Omega_1,\Omega_2)+iJ_{\delta t^2}(\Omega,H).\\
 &\nonumber
\end{eqnarray}
With simple calculations one can now evaluate $\mathrm{Var}_1^{[\psi]}(\delta t)\equiv\mathrm{Var}_1^{[\psi]}(U(\delta t)^{\dag}U(\delta t))$ on a generic state $|\psi\rangle$. It results
\begin{equation}\begin{split}\label{varudu}
\mathrm{Var}_1^{[\psi]}(\delta t)&=\langle\psi|\Big(U(\delta t)^{\dag}U(\delta t)\Big)^2|\psi\rangle-\langle\psi|U(\delta t)^{\dag}U(\delta t)|\psi\rangle^2\\
&=\langle\psi|I_{\delta t^2}^{(2)}(t)|\psi\rangle-\langle\psi|I_{\delta t}^{(1)}(t)|\psi\rangle^2.
\end{split}\end{equation}
The nonthermal drift entropy flux of the $k$-th drift depends on $\kappa_k=-\frac{\mathrm{Var}_1^{[\psi_k]}(U(\delta t)^{\dag}U(\delta t))}{||U(\delta t)|\psi_k\rangle||^4}$ and is then generated, up to second order in $\delta t$, only by the quantum fluctuations of the operator $I_{\delta t}^{(1)}(t)$ on the trajectory state. It is worth stressing that in Eq.~(\ref{varudu}) the first non vanishing contribution is of order $\delta t^2$, while the term $||U(\delta t)|\psi_k\rangle||^4$ has also contributions of order $\delta t^0$ and $\delta t$: This means that the drift entropy flux can be made vanishingly small by choosing a very high measurement rate, such that all terms of order $\delta t^2$ become negligible. Note however that there are at least two lower bounds to $\delta t$: one is given by the measurement speed of the experimental apparatus, which is not infinite. The other one is given by the requirement that the dynamics is not frozen due to Zeno effect: therefore $\delta t$ has to be always greater than the Zeno time of the total system. These two limitations, in some cases, may lead to a nonvanishing drift entropy flux, which is then a quantity of real physical interest.


\begin{thebibliography}{37}
 \bibitem{jarzrev} C. Jarzynski, Annu. Rev. Condens. Matter Phys. \textbf{2}, 329 (2011).
 \bibitem{esporev} M. Esposito, U. Harbola and S. Mukamel, Rev. Mod. Phys. \textbf{81}, 1665 (2009).
 \bibitem{jarz} C. Jarzynski, Phys. Rev. Lett. \textbf{78}, 2690 (1997).
 \bibitem{crooks} G. E. Crooks, Phys. Rev. E \textbf{60}, 2721 (1999).
 \bibitem{espo} M. Esposito and C. Van den Broeck, Phys. Rev. Lett. \textbf{104}, 090601 (2010).
 \bibitem{saga} T. Sagawa and M. Ueda, Phys. Rev. Lett. \textbf{104}, 090602 (2010).
 \bibitem{garcia} R. Garc\'{i}a-Garc\'{i}a, Phys. Rev. E \textbf{86}, 031117 (2012).
 \bibitem{sciencejarz} J. Liphardt, S. Dumont, S. B. Smith, I. Tinoco Jr. and C. Bustamante, Science \textbf{296}, 1832 (2002).
 \bibitem{seif1} S. Schuler, T. Speck, C. Tietz, J. Wrachtrup and U. Seifert, Phys. Rev. Lett. \textbf{94}, 180602 (2005).
 \bibitem{seif2} C. Tietz, S. Schuler, T. Speck, U. Seifert and J. Wrachtrup, Phys. Rev. Lett. \textbf{97}, 050602 (2006).
 \bibitem{genjarz} S. Toyabe, T. Sagawa, M. Ueda, E. Muneyuki and M. Sano, Nature Phys. \textbf{6}, 988 (2010).
 \bibitem{hanggirev} M. Campisi, P. H\"{a}nggi and P. Talkner, Rev. Mod. Phys \textbf{83}, 771 (2011).
 \bibitem{hanggi1} M. Campisi, P. Talkner and P. H\"{a}nggi, Phys. Rev. Lett. \textbf{102}, 210401 (2009).
 \bibitem{muka} M. Esposito and S. Mukamel, Phys. Rev. E \textbf{73}, 046129 (2006).
 \bibitem{kawa} T. Kawamoto and N. Hatano, Phys. Rev. E \textbf{84}, 031116 (2011).
 \bibitem{kafri} D. Kafri and S. Deffner, Phys. Rev. A \textbf{86}, 044302 (2012).
 \bibitem{lutz} S. Deffner, M. Brunner and E. Lutz, Europhys. Lett. \textbf{94}, 30001 (2011).
 \bibitem{liu} F. Liu, Phys. Rev. E \textbf{86}, 010103(R) (2012).
 \bibitem{seif} U. Seifert, Phys. Rev. Lett. \textbf{95}, 040602 (2005).
 \bibitem{deff} S. Deffner and E. Lutz, Phys. Rev. Lett. \textbf{107}, 140404 (2011).
 \bibitem{lnbm} B. Leggio, A. Napoli, H.-P. Breuer and A. Messina, Phys. Rev. E \textbf{87}, 032113 (2013).
 \bibitem{wise1} H. M. Wiseman, Phys. Rev. A \textbf{49}, 2133 (1994).
 \bibitem{wise2} H. M. Wiseman, Quant. Semiclass. Opt. \textbf{8}, 205 (1996).
 \bibitem{hanggi} M. Campisi, P. Talkner and P. H\"{a}nggi, Phys. Rev. Lett. \textbf{105}, 140601 (2010).


\bibitem{wiseman} H. M. Wiseman and G. J. Milburn,
Phys. Rev. A \textbf{47}, 1652 (1993).

\bibitem{carmichael} H. Carmichael, {\textit{An open systems approach to
Quantum optics}}, Lecture Notes in Physics, Vol. m18 (Springer, Berlin, 1993).

 \bibitem{dal} J. Dalibard, Y. Castin and K. M\o{}lmer,  Phys. Rev. Lett. \textbf{68},
 580 (1992).

\bibitem{zoller} R. Dum, P. Zoller and H. Ritsch,
Phys. Rev. A \textbf{45}, 4879 (1992).

 \bibitem{plen} M. B. Plenio and P. L. Knight, Rev. Mod. Phys. \textbf{70}, 101
 (1998).

 \bibitem{breub} H.-P. Breuer and F. Petruccione, \textit{The Theory of Open Quantum Systems} (Oxford University Press, Oxford, 2002).
 \bibitem{mahler} G. Mahler and V. A. Weberuss, \textit{Quantum Networks: Dynamics of Open Nanostructures} (Springer, Berlin, 1998).
 \bibitem{plastina} S. Lorenzo, F. Plastina and M. Paternostro, Phys. Rev. A \textbf{88}, 020102(R) (2013).
 \bibitem{lebo} J. L. Lebowitz and H. Spohn, J. Stat. Phys. \textbf{95}, 333 (1999).
 \bibitem{hpvar} H.-P. Breuer and F. Petruccione, Phys. Rev. A \textbf{54}, 1146 (1996).
 \bibitem{pek} F. W. J. Hekking and J. P. Pekola, Phys. Rev. Lett. \textbf{111}, 093602 (2013).
 \bibitem{horow} J. M. Horowitz and J. M. R. Parrondo, New J. Phys. \textbf{15}, 085028 (2013).
 \bibitem{therm} N. Linden, S. Popescu, A. J. Short and A. Winter, Phys. Rev. E \textbf{79}, 061103 (2009).
\end{thebibliography}
\end{document}